\documentclass[amsmath,amssymb,aps,prb,reprint,superscriptaddress]{revtex4-2}
\usepackage[utf8]{inputenc}
\usepackage{graphicx}
\usepackage{amsmath}
\usepackage{amssymb}
\usepackage{float}
\usepackage{booktabs}
\usepackage{tabularx}
\usepackage{array}
\usepackage{longtable}
\usepackage{multirow}
\usepackage{longtable}
\usepackage{adjustbox}
\usepackage{booktabs}
\usepackage{supertabular}
\usepackage{diagbox}
\usepackage{setspace}
\usepackage{xcolor}
\usepackage{tikz}
\usepackage{dcolumn}
\usepackage{bm}
\usepackage{xcolor}

\usetikzlibrary{positioning} 

\newcommand{\fatq}{\mathbf{q}}
\newcommand{\fatQ}{\mathbf{Q}}

\newcommand{\pgnfigure}[2]{\begin{figure}[h]\includegraphics[width=7.4cm]
{#1}\caption{\label{#1}#2}\end{figure}}

\begin{document}

\title{Spin-density wave of ferrimagnetic building blocks masking the ferromagnetic quantum-critical point in NbFe$_{2}$}

\author{T. Poulis}
\affiliation{Department of Physics, Royal Holloway, University of London, Egham TW20 0EX, United Kingdom}
\author{G. Mani}
\affiliation{Department of Physics, Royal Holloway, University of London, Egham TW20 0EX, United Kingdom}
\author{J. Sturt}
\affiliation{Department of Physics, Royal Holloway, University of London, Egham TW20 0EX, United Kingdom}
\author{W. J. Duncan}
\affiliation{Department of Physics, Royal Holloway, University of London, Egham TW20 0EX, United Kingdom}
\author{H. Thoma}
\affiliation{Institute of Crystallography, RWTH University, Aachen 52066, Germany}
\author{V. Hutanu}
\affiliation{Institute of Crystallography, RWTH University, Aachen 52066, Germany}
\author{B. Ouladdiaf}
\affiliation{Institute Laue Langevin, Grenoble 38042 - CS 20156, France}
\author{I. Kibalin}
\affiliation{Institute Laue Langevin, Grenoble 38042 - CS 20156, France}
\author{M. H. Lemee}
\affiliation{Institute Laue Langevin, Grenoble 38042 - CS 20156, France}
\author{P. Manuel}
\affiliation{Isis Neutron Source, STFC Rutherford Appleton Laboratory, Didcot OX11 0QX, United Kingdom}
\author{A. Neubauer}
\affiliation{Physik Department E21, Technische Universität München, 85748 Garching, Germany}
\author{C. Pfleiderer}
\affiliation{Physik Department E21, Technische Universität München, 85748 Garching, Germany}
\author{F. M. Grosche}
\affiliation{Cavendish Laboratory, University of Cambridge, Cambridge CB3 0HE, United Kingdom}
\author{P. G. Niklowitz}
\affiliation{Department of Physics, Royal Holloway, University of London, Egham TW20 0EX, United Kingdom}

\date{\today}

\begin{abstract}

In the metallic magnet NbFe$_2$, the low temperature threshold of ferromagnetism can be investigated by varying the Fe concentration within a narrow homogeneity range. NbFe$_2$ is one of a number of compounds where modulated order is found to mask the ferromagnetic quantum critical point. However, here we report the rare case where the masking modulated magnetic order has been fully refined. Spherical neutron polarimetry and high-intensity single-crystal neutron diffraction reveal the first case of a longitudinal spin-density wave masking the ferromagnetic quantum critical point. The spin-density wave is characterised by a large-wavelength incommensurate modulation of its low average moment. It is formed from ferrimagnetic building blocks with antiparallel ferromagnetic sheets. The existence of ferromagnetic sheets and compensation of the magnetisation only over mesoscopic length scales show local similarity between the spin-density wave and the ferromagnetic parent phase and indicate the spin-density wave's unconventional nature as emerging from underlying ferromagnetic quantum criticality.


\end{abstract}

\maketitle

\section{Introduction}

The exploration of ferromagnetic (FM) quantum phase transitions in metals has motivated numerous theoretical and experimental studies \cite{bra16a}, which have led to the discovery of non-Fermi liquid states \cite{uhl04a,nik04d} and of unconventional superconductivity (e.g. \cite{pfl02a,sax00a,ali10a}). The underlying question, however, whether an FM quantum critical point (QCP) can exist in clean band magnets, remains controversial. Fundamental considerations \cite{pfl09a,voj99a,chu04a} suggest that the FM QCP is avoided in clean systems by one of two scenarios: either the transition into the FM state becomes discontinuous (first order), or the nature of the low temperature ordered state changes altogether, for instance into nematic or long-wavelength spin density wave (SDW) order \cite{voj99a,chu04a}. Examples for the first scenario, include ZrZn$_2$ \cite{uhl04a,kab13a}, Ni$_3$Al \cite{nik04d} and UGe$_2$ \cite{pfl02a,tau10a}. Examples for the transition into a modulated state include: (i) the masking of the field-tuned quantum-critical end point of the continuous metamagnetic transition of Sr$_3$Ru$_2$O$_7$ by two SDW orders \cite{les15a} with enhanced spin fluctuations in the ordered states \cite{les21a} and the field-induced suppression of FM and replacement with a modulated fan state in the local moment system PrPtAl \cite{abd15a,one21a}, (ii) modulated order in the heavy-fermion system YbRh$_2$Si$_2$ that is adjacent to FM under Co-doping \cite{lau13a,ham19a}, features a high Wilson ratio \cite{geg05a} and the evolution of FM fluctuations into long-wavelength SDW fluctuations upon cooling \cite{sto12a} and hosts a candidate for odd-parity superconductivity boosted by an electro-nuclear SDW \cite{sch16a,ngu21a,lev26a}, (iii) and at the border of pressure-tuned FM the appearance of modulated magnetic order in CeRuPO \cite{len15a} and helical magnetic order terminating in superconductivity in MnP \cite{che15a,mat16a,kha16a,dis23a}, while pressure-induced short-range FM clusters are reported at the border of long-range FM in LaCrGe$_3$ \cite{kal17a,gati2021formation}. 

\pgnfigure{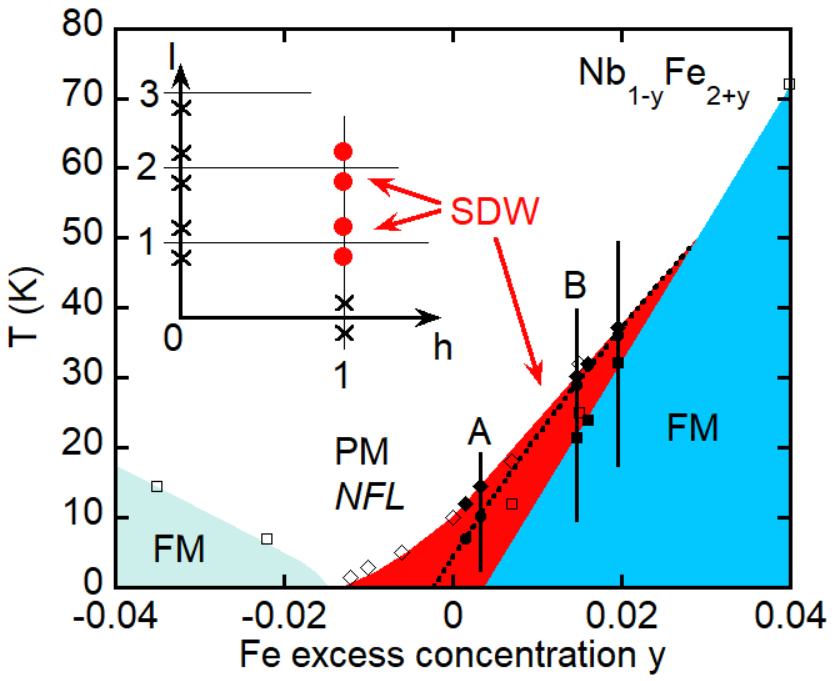}{Phase diagram of Nb$_{1-y}$Fe$_{2+y}$ with results for bulk $T_{\rm C}$ (squares) and $T_{\rm N}$ (diamonds) from previous single-crystal (filled symbols) \cite{niklowitz2019ultrasmall} and polycrystal (empty symbols) \cite{mor09a} studies. Vertical solid lines indicate the $T$ range of previous cold neutron diffraction measurements. 'A' and 'B' denote samples also studied with thermal neutron diffraction reported here. Of the two ferromagnetic (FM) phases, the one on the more Fe-rich side is separated from the paramagnetic (PM) state by a spin-density wave (SDW) at low temperatures, where non-Fermi liquid (NFL) behaviour is found as well. $T_{\rm 0}$, the FM phase boundary buried by the SDW phase (dashed line) is an extrapolation of $T_{\rm 0}$ values (circles) measured or calculated for the single crystals. The inset shows the relevant reciprocal-space region, which was accessible during previous neutron diffraction experiments. Circles show the presence and crosses the absence of SDW peaks.}

An incommensurate SDW masking the border of itinerant FM at zero temperature, field, and pressure is found in Nb$_{1-y}$Fe$_{2+y}$. The border of FM is located near stoichiometric NbFe$_2$ \cite{shi87a}, where it is masked by modulated order. FM order can be induced at low temperature by growing Fe-rich Nb$_{1-y}$Fe$_{2+y}$ with $y$ as small as 1\% (Fig. \ref{figure_1} \cite{mor09a}). The SDW nature of the modulated order at the border of FM was indicated by NMR \cite{yam88a}, ESR, $\mu$SR, and M\"o{\ss}bauer spectroscopy \cite{rau15a}. Neutron scattering  revealed a long-wavelength SDW with associated ordering wavevector $\fatq_{\rm SDW}=$(0,0,$l_{\rm SDW}$). $l_{\rm SDW}$ is of the order $0.1$ r.l.u but found to depend substantially on $y$ and $T$, decreasing as the FM state is approached but staying finite at the first-order SDW-FM phase transition \cite{niklowitz2019ultrasmall}. Non-Fermi liquid forms of resistivity and low temperature heat capacity have been observed in slightly Nb-rich NbFe$_2$ in the vicinity of the SDW phase. \cite{bra08a} The SDW's structure factor and therefore its microscopic details within the NbFe$_2$ unit cell including the effective moment size and orientation at the individual magnetic sites have been unknown so far. 

Here we present the full magnetic structure of the spin-density-wave phase. Spherical neutron polarimetry (SNP) excluded spiral order and transverse SDW order as reported for other compounds \cite{les15a, abd15a}. High-intensity single-crystal neutron diffraction (ND) in particular has allowed full refinement despite a low average moment of 0.039~$\mu_B$/atom. A longitudinal spin-density wave with a large-wavelength incommensurate modulation along the magnetic easy $c$-axis is formed from ferrimagnetic unit-cell building blocks with antiparallel ferromagnetic sheets stacked along the $c$-axis. In particular, this excludes the possibility of a compensation of the magnetisation within the NbFe$_2$ unit cell. The compensation of the magnetisation only over length scales of 50 - 90 \AA\ indicates the spin-density wave's unconventional nature as emerging from underlying ferromagnetic quantum criticality.

\section{Experimental Details}

Large single crystals of C14 Laves phase NbFe$_2$ (lattice constants $a\approx4.84$\,\AA\ and $c\approx7.89$\,\AA) with compositions chosen across the iron-rich side of the homogeneity range have been grown in a UHV-compatible optical floating zone furnace from polycrystals prepared by induction melting \cite{neu11a}. The single crystals have been characterised extensively by resistivity, susceptibility, and magnetisation measurements, as well as by x-ray diffraction and neutron depolarization \cite{pfl10a,neu11b,dun11a}, the latter showing homogeneity in structure and chemical composition. 

Here, two samples that were already used to measure the SDW ordering wavevector \cite{niklowitz2019ultrasmall} have been studied: (i) sample A is almost stoichiometric ($y = +0.003$) with $T_{\mbox{N}}$ at $15.3$\,K (revisited from the value reported in \cite{niklowitz2019ultrasmall} based on more detailed low-$T$ measurements as described in Supplement) and an SDW ordering vector $q_{SDW}=$(0 0 0.156); (ii) sample B is slightly Fe-rich ($y = +0.015$) with $T_{\mbox{N}}$ and $T_{\mbox{C}}$ at $32.5$\,K and $19$\,K, respectively, and $q_{SDW}=$(0 0 0.10).

The set of polycrystals and single crystals (including the samples of this study) \cite{mor09a,dun11a} on which the phase diagram in Fig. \ref{figure_1} is based have residual resistivities in the range of 5 to 19 $\mu\Omega$cm. According to theoretical studies of the disorder dependence of FM quantum phase transitions \cite{belitz1999first,san14a} this would place them at the lower end of the intermediate regime (several to hundreds of $\mu\Omega$cm) for which a FM QCP is predicted. However, those predictions are rough order-of-magnitude estimates, which means that the observation of the the phase diagram in Fig. \ref{figure_1} that matches the prediction for the clean limit does not represent a clear contradiction with theory.

For the neutron scattering experiments the samples were mounted on Al holders. The nature of magnetic order was determined with spherical neutron polarimetry (SNP) using Cryopad at the polarized hot neutron diffractometer POLI (MLZ) \cite{mlz15a}. The beamline set-up included polarized \(^3\text{He}\) spin filters as both polarizers and analyzers, and a Cu crystal monochromator set to select a neutron wavelength of \(\lambda = 0.9 \, \text{\AA}\). The sample was mounted on a standard closed-cycle cryostat.

The full magnetic structure was obtained through unpolarised single-crystal neutron diffraction (ND) at four-circle diffractometer D10+ (ILL) \cite{nik23a}. The monochromator was set to \(\lambda = 2.363 \, \text{\AA}\). A pyrolytic graphite (PG) filter suppressed higher-order harmonics. An analyser to reduce the quasielastic background was essential to detect the SDW Bragg peak signals that were collected with a single-counter \(^3\text{He}\) detector. The sample was cooled with a Joule-Thomson displex. A few magnetic peaks have additionally been measured with time-of-flight (TOF) spectrometer WISH at ISIS \cite{nik26a_sup}.

\section{Results}

SNP was done on slightly Fe-rich Sample B at $T=25\,$K in the SDW state. Background measurements at $T=50\,$K confirmed the PM phase. The full set of 36 channels was obtained at the two strong magnetic Bragg peaks, Q=(2 $\overline{2}$  $\overline{1.1})$ and Q=(2 $\overline{2}$ $\overline{3.1})$. Scattering factors at 25\,K have been obtained from two independent analysis methods: a numerical method (Table \ref{scattering_factors}) and an analytical method (see Supplement \cite{nik26a_sup}, which agree within the error.  

\begin{table}[H]
\centering
\caption{Scattering factors of nuclear ($N$) and magnetic ($M$) scattering obtained from least-squares fitting of the results at 25\,K for all 36 SNP channels. The coordinates are defined as $x\|\fatQ$, $y\perp x$ and in the $(h0l)$ scattering plane, and $z\perp x,y$.}
\label{scattering_factors}
\renewcommand{\arraystretch}{2} 
\setlength{\tabcolsep}{4pt} 
\resizebox{\columnwidth}{!}{ 
\begin{tabular}{c c c} 
\textbf{\large Scattering Factors} & \textbf{\large Q=$(2$ $\overline{2} $ $\overline{1.1}$)}   & \textbf{\large Q=$(2$ $\overline{2}$  $\overline{3.1}$)}   \\ 
\textbf{\large $NN^{*}$}              & \textbf{\large $0.025 \pm 0.011$}     & \textbf{\large $0.002 \pm 0.000$}    \\ 
\textbf{\large $M_yM^{*}_y$}             & \textbf{\large $7.559 \pm 0.455$}      & \textbf{\large $2.450 \pm 0.176$}      \\ 
\textbf{\large $M_zM^{*}_z$}             & \textbf{\large $0.019 \pm 0.018$}     & \textbf{\large $0.014 \pm 0.014$}       \\ 
\textbf{\large $i(M_{\perp}\times M^{*}_{\perp})$}  & \textbf{\large $-0.340 \pm 0.020$} & \textbf{\large $0.090 \pm 0.100$}     \\ 
\textbf{\large $Re(M_yM^{*}_z)$}       & \textbf{\large $-0.150 \pm 0.300$}    & \textbf{\large $-0.140 \pm 0.109$} \\ 
\textbf{\large $Re(NM^{*}_y)$}       & \textbf{\large $0.299 \pm 0.139$}      & \textbf{\large $0.014 \pm 0.039$}  \\ 
\textbf{\large $Re(NM^{*}_z)$}         & \textbf{\large $-0.003 \pm 0.019$}    & \textbf{\large $-0.003 \pm 0.006$}    \\ 
\textbf{\large $Im(NM^{*}_y)$}       & \textbf{\large $-0.250 \pm 0.025$} & \textbf{\large $0.150 \pm 0.099$} \\
\textbf{\large $Im(NM_z^{*})$}        & \textbf{\large $0.014 \pm 0.016$}     & \textbf{\large $-0.013 \pm 0.013$}    \\ 
\end{tabular}}
\end{table}

The scattering factors in Table \ref{scattering_factors} reveal that the SDW phase is formed by collinear magnetic moments along the easy $c$ axis. The only positive signals substantially larger than the error range at both investigated magnetic Bragg peaks are found for $M_yM_y^{*}$. In particular, the $M_zM_z^{*}$ signal of $z$ components in the hexagonal $ab$ plane (\ref{scattering_factors}) is within the noise. Given the hexagonal symmetry of the Nb$_{1-y}$Fe$_{2+y}$ crystal structure, magnetic order is often expected to preserve or respect the $ab$-plane symmetry. A zero $M_zM_z^{*}$ signal is only possible if the SDW moments have their spins predominantly oriented on the c axis. This and the previously reported ordering wave vector $\fatq_{\rm SDW}=$(0,0,$l_{\rm SDW}$) \cite{niklowitz2019ultrasmall} means that the SDW in Nb$_{1-y}$Fe$_{2+y}$ is longitudinal. This conclusion is also in agreement with the previously reported absence of magnetic diffraction peaks at $\fatq=($ $0$ $0$ $l\pm l_{\rm SDW})$ with $l=1,2,3$ \cite{niklowitz2019ultrasmall} when taking into account the relevant neutron selection rule. 

Candidates for the SDW's irreducible representations were identified with the SARAh Refine program \cite{wil00a} based on the NbFe$_2$ crystal structure and its SDW propagation vector. Four possible irreducible representations were found for the Fe$_{2a}$ site, six for Fe$_{6h}$, and the same four as for Fe$_{2a}$ for each of the two Nb sites that separate into two distinct magnetic sites (Nb$_1$ and Nb$_2$) within the $P6_3/mmc$ space group. Two further constraints are: (i) the SNP result described above that the SDW is formed of magnetic moments along the $c$ axis, and (ii) the SDW phase is separated from PM by a single critical temperature, which points to a single common irreducible representation for all magnetic sites \cite{wills2001magnetic}. Only two such representations out of the six satisfy these further constraints: one candidate is the $\Gamma_2$ representation where the magnetic moments on symmetry-equivalent sites are ferromagnetically aligned. The other candidate is $\Gamma_4$ with antiferromagnetic alignment between symmetry-equivalent sites. Both representations include, for the Fe atoms on the $6h$ Wyckoff position the potential for in-plane ($ab$) components. These in-plane components would cause three Fe atoms of kagome triangle to orient in 'all-in' or 'all-out' configurations. Those do not contribute to the unit-cell's net magnetisation, which for $\Gamma_2$ and $\Gamma_4$ can therefore only be collinear with the easy $c$ axis.

Unpolarised ND was collected on almost stoichiometric ($y = +0.003$) Sample A at a $T=4\,$K in the SDW state. SDW peaks at 104 different positions falling into 23 sets of symmetrically independent reflections were measured to enable full structure refinement. The magnetic peaks were fitted with Gaussian functions. Magnetic intensities were corrected for two effects (see Supplement): (i) amplitude reduction of strong nuclear peaks by detector saturation and (ii) a single-counter-detector induced reduction of the peak width beyond the $q$ dependent instrument-resolution.

\begin{figure}[h!]
    \centering
    \includegraphics[width=\columnwidth]{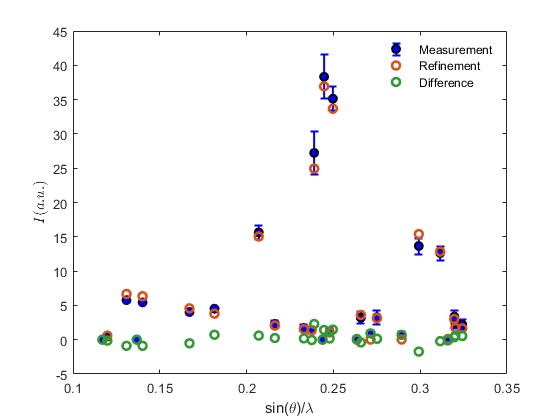}
    \caption{Refinement of SDW Bragg peaks from unpolarised ND by the $\Gamma_2$ representation that leads to the best agreement (see fit parameters in Table \ref{refinement_result_pars}). }
    \label{refinement}
\end{figure}

\begin{figure}[h!]
    \centering
    \begin{tikzpicture}
        \node[inner sep=0pt] (big) at (-12,0)
            {\includegraphics[width=0.3\linewidth,keepaspectratio]{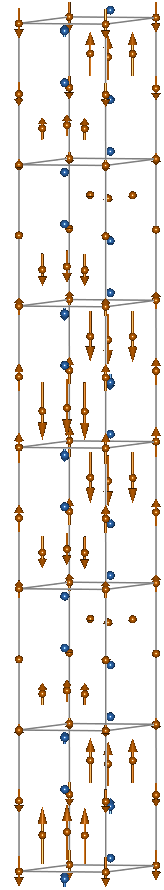}};

        \node[inner sep=0pt] (zoom) at (-7.5,0)
            {
                \begin{minipage}{0.7\linewidth}
                    \includegraphics[width=\linewidth,keepaspectratio]{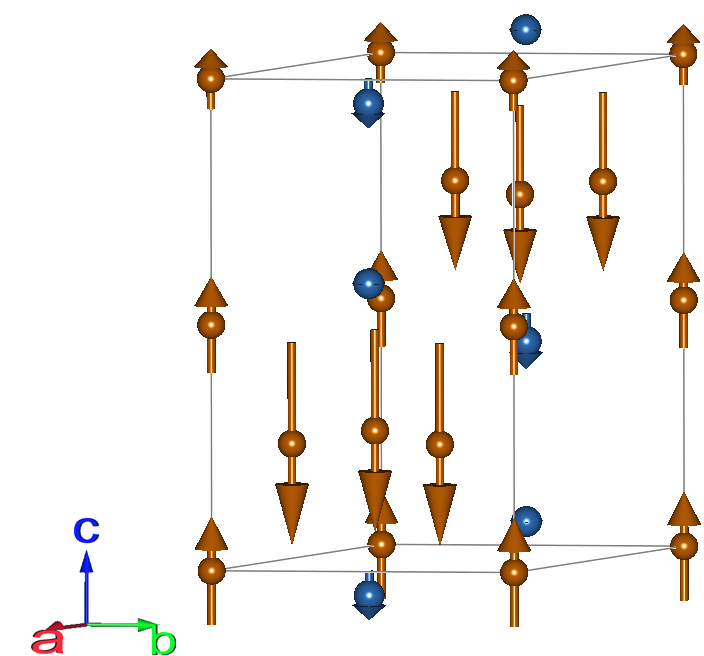}
                \end{minipage}
            };

        \path[use as bounding box]
            (big.north west) rectangle (zoom.south east);

        \draw[thick,->] (-10.7,0.65) -- (-9.2,0.65);
       
    \end{tikzpicture}
    \caption{Six-unit-cell SDW structure with highlighted region (left) and zoomed-in single-unit-cell model (right), including Cartesian axes, of the ferrimagnetic building block of the incommensurate long-wavelength SDW structure of $NbFe_2$ from refinement of unpolarised ND data. Fe atoms (orange) and Nb atoms (blue) each form ferrimagnetic sublattices.}
    \label{fig:SDW_structure}
\end{figure}

Corrected SDW peak intensities were used for structural refinement with FullProf. The $\Gamma_2$ representation that meets all constraints discussed above from crystal symmetry, SNP results and the phase diagram can be brought into excellent agreement with the data, as shown in Fig. \ref{refinement}. All basis vectors permitted by the $\Gamma_2$ representation were refined simultaneously and the goodness of fit parameters are presented in Table \ref{refinement_result_pars}. Despite the relatively weak magnetic intensities and moments, refinement results with R values $<15\%$ are quite robust. Fig.\ref{fig:SDW_structure} depicts the magnetic-moment arrangement in a crystal-lattice unit cell that acts as a building block for the SDW structure with its longer-range incommensurate modulation. Within such a building block the $\Gamma_2$ representation is characterised by parallel arrangement of magnetic moments at a given type of magnetic site. This creates ferromagnetic layers parallel to the $ab$ plane. Neighbouring Fe layers are aligned anti-parallel with different moment-sizes, creating a ferrimagnetic arrangement within the building block. The Nb sublattice also carries magnetic order with the net moment pointing in the same direction as the net moment of the Fe lattice. Models based on other representations including those where moments exclusively have $ab$ plane components were also tested. None provided a satisfactory fit to the data, with refinement of $\Gamma_4$ as the next best representation leading to at least an order of magnitude larger goodness-of-fit values. 

 \begin{table}[h!]
    \centering
    \setstretch{1}
    \caption{Refined amplitudes of the spin--density--wave (SDW) magnetic moments (in $\mu_B$) along the crystallographic $a$, $b$, and $c$ axes. The amplitude of the net unit-cell magnetic moment is $m_{\mathrm{amp}}=\frac{1}{n}\lvert\sum_{i=1}^{n} S_i\rvert$. The average magnitude of the net unit-cell magnetic moment is
$\bar{m}_{\mathrm{av}} = \frac{1}{n\,N_{\mathrm{cell}}}
\sum_{\ell=1}^{N_{\mathrm{cell}}}
\biggl|
\sum_{j=1}^{n} S_{j}
\biggr|$. Uncertainties reflect the propagation of errors from the FullProf basis--vector refinements. Goodness--of--fit parameters are listed below.} 
    \label{refinement_result_pars}
    \resizebox{\columnwidth}{!}{ 
    \begin{tabular}{lccc}
        \hline
        Site & $S_a$ ($\mu_B$) & $S_b$ ($\mu_B$) & $S_c$ ($\mu_B$) \\
        \hline
        Fe$_{2a}$ & 0 & 0 & $0.0439 \pm 0.0021$ \\
        Fe$_{6h}$ & 0  & 0 & $-0.0849 \pm 0.0016$ \\
        Nb$_1$    & 0 & 0 & $-0.0135 \pm 0.0202$ \\
        Nb$_2$    & 0 & 0 & $-0.0097 \pm 0.0066$ \\

        \hline
        $m_{\mathrm{amp}}$ ($\mu_B$/atom) & $0.0390 \pm 0.0062$ \\
        $\bar{m}_{\mathrm{av}}$ ($\mu_B$/atom) & $0.0249 \pm 0.0062$ \\
        $R_{F^2}$                         & 8.306 \\
        $R_{F^2w}$                        & 13.11 \\
        $R_F$                             & 8.875 \\
        $\chi^2$                          & 1.583 \\
        \hline
    \end{tabular}}
\end{table}

\section{Discussion}

The magnetic-moment amplitudes obtained from the refinement of unpolarised ND intensities (Table \ref{refinement_result_pars}) result in an amplitude of the net unit-cell magnetic moment of $m_{\mathrm{amp}}=\frac{1}{n}|\sum_{i=1}^{n} S_i|=0.0390\pm 0.0062$\,$\mu_{\rm B}$/atom where $n$ is the number of atoms. Due to the long-range modulation of the unit-cell structure the average magnitude of the net unit-cell magnetic moment is $\bar{m}_{\mathrm{av}} = \frac{1}{n\,N_{\mathrm{cell}}}
\sum_{\ell=1}^{N_{\mathrm{cell}}}
\biggl|
\sum_{j=1}^{n} S_{j}
\biggr| = 0.0249\pm 0.0062$\,$\mu_{\rm B}$/atom, where $S_i$ is the magnetic moment of the spin $i$, $n$ the number of atoms in a unit cell and $N_{cell}$ the number of unit cells, in excellent agreement with muon-spin rotation ($\mu$SR) measurements \cite{wil22a} that reported $0.03~\mu_{\rm B}$ per Fe atom. The minimal size of magnetic-moment components in the $ab$ plane agrees with the SNP results reported above. 

The presence of FM sheets in the $ab$ plane has also been seen in TiFe$_2$ \cite{brown1992magnetic}. This is another Laves phase compound whose magnetic order is sensitive to stoichiometry, with the bulk of the magnetic moments parallel to the c-axis and concentrated on the Fe$_{6h}$ iron sheets, which are coupled ferromagnetically. However, in contrast to TiFe$_2$, for NbFe$_2$, the assumption of no magnetic moments on the Fe$_{2a}$ sites leads to poor refinement of the ND data although the magnetic moments on those sites in NbFe$_2$ are very small. For other Laves phase compounds, their absence has been related to topological frustration \cite{brown1992magnetic} although the role of frustration in itinerant magnets tends to be reduced. 

A main consequence of the emergence of the longitudinal incommensurate SDW in NbFe$_2$ is that the formation of the FM QCP is prevented. To the authors' knowledge, this is the first observation of a longitudinal SDW emerging from the border of FM, but fan-like SDW order is reported to mask a magnetic-field tuned QCP in PrPtAl \cite{one21a} and transverse SDW order is reported to mask a magnetic-field tuned metamagnetic quantum-critical endpoint in Sr$_3$Ru$_2$O$_7$.\cite{les15a} Furthermore, helical magnetic order appears to mask a FM QCP in MnP under pressure.\cite{kha16a,mat16a,dis23a} These examples support theories that predict the intrinsic instability of QCPs in clean ferromagnets and their replacement by modulated magnetic phases \cite{bel97a,voj99a,chu04a}. The latter phases are stabilised by offering increased phase space for electronic low-energy particle-hole excitations, a mechanism known as quantum order by disorder \cite{kar12a,gre18a}. NbFe$_2$ as a band magnet hosting this scenario at zero field will allow to test those theories in their most basic form.

The ferrimagnetic stacking of FM layers along the $c$-axis means that there is a finite net magnetisation within the crystal-lattice unit-cell building block - in contrast to, e.g., antiferromagnetic stacking. The local finite net magnetisation in NbFe$_2$ is therefore only compensated over the mesoscopic length scales of the incommensurate SDW modulation of $50-90\,$\AA\ and the SDW mimics FM at microscopic length scales. Similar length scales for magnetisation compensation are observed for for the modulated component in PrPtAl (approx. $30\,$\AA), for Sr$_3$Ru$_2$O$_7$ (approx. $25\,$\AA) and for MnP (approx. $15\,$\AA), which suggests that this could be a generic feature of modulated order emerging at the border of FM. 

The ferrimagnetic moment configuration in the SDW building blocks suggests that the magnetic parent phase of NbFe$_2$ is not FM, but ferrimagnetic as well. Density functional theory studies (DFT) have suggested ferrimagnetism as a possible ground state of NbFe$_2$ among a number of candidate magnetic ground states with very small energy differences between them \cite{sub10a,tom10a,nea11a}. The calculations overestimate the magnetic site moment magnitudes, as is often the case for systems with enhanced magnetic fluctuations. Compton scattering results on the Fe-rich side of the phase diagram have also been analyzed by assuming ferrimagnetism as the ground state \cite{hay12a}. Because of the location of the stopping site, $\mu$SR data \cite{wil22a} could not distinguish between FM and ferrimagnetic order. M{\"o}{\ss}bauer results have been interpreted as indication of FM order \cite{rau15a}.

If one assumed ferrimagnetic order for the parent phase, it would be intuitive to imagine it to simply be the SDW order without the long-wavelength modulation, which would have a net magnetisation of approximately 0.0390\,$\mu_{\rm B}$/atom. If one assumes FM order for the parent state, the natural guess would be a parallel alignment of all moments listed in Table \ref{refinement_result_pars}. This results in an average magnetisation of 0.0536\,$\mu_{\rm B}$/atom. Magnetisation measurements in the parent phase in Fe-rich samples  phase~\cite{friedemann2013ordinary} detect an average magnetisation of approximately 0.06~$\mu_B$ per atom. While this comparison leans more to the parent phase being ferromagnetic, a conclusive answer will likely require a direct observation of the magnetic parent-phase structure by polarised neutron diffraction. 

\section{Conclusion}

This work shows that the emerging modulated structure of the itinerant ferromagnet NbFe$_2$ is an incommensurate long-wavelength SDW formed from ferrimagnetic blocks of FM planes stacked along the easy $c$ axis. The existence of ferromagnetic sheets and compensation of the magnetisation only over substantial length scales indicates the spin-density wave's unconventional nature as emerging from underlying ferromagnetic quantum criticality.

\begin{acknowledgments}

This work is based upon experiments performed at D10+ of Institut Laue-Langevin (ILL), Grenoble, France \cite{nik23a} and at the POLI instrument operated by JCNS/ Forschungszentrum J\"{u}lich GmbH and RWTH Aachen University at the Heinz Maier-Leibnitz Zentrum (MLZ), Garching, Germany. Further experiments at the WISH beamline of the ISIS Neutron and Muon Source were supported by beamtime allocation from the Science and Technology Facilities Council. \cite{nik23b}. The instrument POLI \cite{mlz15a} received funding from the German Federal Ministry of Research, Technology and Space (BMFTR, formerly BMBF) in the framework of the ErUM-Pro programme (grant numbers 05K10PA2, 05K13PA3). The authors gratefully acknowledge the financial support provided by the ILL to perform the neutron scattering measurements there and by JCNS/ Forschungszentrum J\"{u}lich GmbH to perform the neutron scattering measurements at MLZ. The authors acknowledge support by the EPSRC through grant EP/K012894/1. The authors thank D. Voneshen for assistance with Mantid Project software and N. A. Katcho and E. Canonero for helpful discussions. 

 \end{acknowledgments}

\bibliographystyle{apsrev4-2}
\bibliography{references,refs_pgn}  

\end{document}